\documentclass{emulateapj}

\usepackage{lscape}

\begin{document}

\title{
Crystalline Silicate Emission in the Protostellar Binary Serpens--SVS20}

\author{David R. Ciardi}
\affil{Michelson Science Center/Caltech\\
 770 South Wilson Avenue, M/S 100-22\\
Pasadena, CA 91125}
\email{ciardi@ipac.caltech.edu}

\author{Charles M. Telesco, Christopher Packham, \\
Cynthia G\'omezMartin, James T. Radomski}
\affil{University of Florida, Department of Astronomy\\
211 Space Sciences Building, Gainesville, FL 32611}

\author{James M. De Buizer}
\affil{Gemini Observatory\\
Casilla 603, La Serena, Chile }

\author{Chris J. Phillips}
\affil{Australia Telescope National Facility, CSIRO\\
P.O. Box 76, Epping NSW 1710, Austrailia}

\author{David E. Harker}
\affil{University of California, San Diego, CASS\\
9500 Gilman Dr., Dept. 0424, La Jolla, CA 92093-0424}



\slugcomment{Accepted for publication in The Astrophysical Journal}

\begin{abstract}

We present spatially resolved mid-infrared spectroscopy of the class
I/flat-spectrum protostellar binary system SVS20 in the Serpens cloud core.
The spectra were obtained with the mid-infrared instrument T-ReCS on
Gemini-South. SVS20-South, the more luminous of the two sources, exhibits a
mid-infrared emission spectrum peaking near 11.3 \micron, while SVS20-North
exhibits a shallow amorphous silicate absorption spectrum with a peak optical
depth of $\tau \sim 0.3$.  After removal of the the line-of-sight extinction
by the molecular common envelope, the ``protostar-only'' spectra are found to
be dominated by strong amorphous olivine emission peaking near 10 \micron. We
also find evidence for emission from crystalline forsterite and enstatite
associated with both SVS20-S and SVS20-N. The presence of crystalline silicate
in such a young binary system indicates that the grain processing found in
more evolved HAeBe and T Tauri pre-main sequence stars likely begins at a
relatively young evolutionary stage, while mass accretion is still ongoing.

\end{abstract}

\keywords{infrared: ISM --- infrared: stars --- ISM: individual (Serpens,
SVS20, SMM6) stars: formation --- stars: pre--main-sequence}

\section{Introduction}

Silicates found in the general interstellar medium are amorphous dust grains,
while in T Tauri and HAeBe pre-main sequence stars \citep[e.g.,][]{meeus03,
prz03, vanBoekel03, ks05}, in debris disks surrounding main sequence stars
\citep[e.g., $\beta$ Pic \& HD 145263;][]{tk91, knacke93, honda04}, and in our
own solar system \citep[e.g., Comet C/1996 Q1;][]{harker99}, we observe both
amorphous and crystalline silicate grains. This distinction between the dust
in the ISM and dust in stellar disks implies a transformation of the dust
during the early phases of stellar and disk evolution when planets form. Thus,
it is reasonable that probing the evolution of the silicate grains may provide
insight into the concurrent planetary evolution.

The data illuminating the evolution of the dust are still sparse and, at
times, contradictory. For example, some young stellar objects contain larger
amorphous silicate grains indicative of grain growth, while other young
stellar objects of similar evolutionary status have signs of crystalline
silicate grains, with no simple picture indicating at what stage in the
evolutionary sequence significant grain growth and crystallization take place
\citep[e.g.,][]{meeus03, prz03, vanBoekel03}. While a general evolutionary
sequence of dust processing seems to be emerging \citep[e.g.,][]{ks05}, other
significant issues, such as the possible dependence of the grain evolution on
the source luminosity and temperature, and the effect of nearby companions in
multiple star systems, are not yet understood.

As part of our binary protostellar program \citep[see also][]{crd03}, we
observed the mid-infrared emission of the protostellar binary SVS20
\citep{svs76}. SVS20 is a member of the young protostellar cluster \citep[age
$\sim 10^5$ yr;][]{kaas04} located in the Serpens cloud core \citep[$d\simeq
250$ pc,][]{scb96}. SVS20 is a sub-mm source \citep[SMM6;][]{ced93} and was
first identified as a binary system by \citet{ellh87}. The components
(SVS20-South and SVS20-North) are separated by $1\farcs5$ ($\sim 375$ AU) at a
position angle of $9\fdg 9$. The two sources have similar infrared spectral
energy distributions with spectral indices appropriate for class I/flat
spectrum protostars \citep{haisch02}. From the 3.1 \micron\ ice absorption
feature, \citet{el87} determined that the binary components share a common
envelope with an extinction of A$_{\rm V} \approx 14$ mag. In addition,
near-infrared polarization observations are centro-symmetric about SVS20
indicating that SVS20 has partially evacuated the cavity immediately
surrounding the binary system, potentially offering a more direct view of the
central protostars \citep{hwk97}.

ISOCAM CVF observations of SVS20, with an angular resolution of 6\arcsec\
pixel$^{-1}$, indicated amorphous silicate in both absorption and emission
\citep{alexander03}, but the ISO observations did not resolve SVS20-S and
SVS20-N.  We present spatially resolved $8-13\ \mu$m spectroscopy of SVS20.
The spectrum of SVS20-N is dominated by amorphous silicate absorption, while
the spectrum of SVS20-S exhibits strong crystalline silicate emission. The
clear differences in the spectra indicate differing levels of dust processing
for the two components of the binary system.

\section{Observations and Data Reduction}

\subsection{Spectroscopy}

Mid-infrared ($8-13\ \mu$m) spectroscopic observations of SVS20 were made on
2003 October 05 (UT) using the the Thermal Region Camera and Spectrograph
\citep[T-ReCS; ][]{telesco98} on the Gemini South 8 m telescope. T-ReCS
utilizes a $320 \times 240$ pixel Si:As blocked impurity band detector, with a
spatial scale of $0\farcs089$ pixel$^{-1}$ and a field of view of $28\farcs8
\times 21\farcs6$.  The observations utilized the low-resolution grating ($R
\sim 111$ at $\lambda_c = 10.5\ \mu$m) and a $0\farcs72$ slit. The N-band
filter ($\lambda_o = 10.36\mu {\rm m}, \Delta\lambda=5.2\mu {\rm m}$) served
as the blocking filter.  The spectral dispersion (along the 320 pixel-axis) is
0.022045 $\mu$m pix$^{-1}$.  T-ReCS was rotated so that both components were
positioned in the slit simultaneously. We used a standard 15\arcsec\
north-south chop-nod sequence, with an on-source integration time of 150 s.
Standard NOAO IRAF packages were used to reduce the data and extract the
spectra. HD 187642 (A7V, F$_\nu[9.8\mu{\rm m}]=33\pm1$
Jy\footnote{http://www.gemini.edu/sciops/instruments/miri/filters/stdfluxes\_zero.txt})
was observed for telluric line removal and flux calibration. The zero point of
the wavelength scale was set using the deepest point of the telluric ozone
feature at 9.495 $\mu$m. The signal-to-noise of the spectra is S/N $\approx
50-70$, where the noise has been estimated from the pixel-to-pixel variations
in the spectra.

\subsection{Photometry}

Supplemental photometry was acquired on SVS20 with T-ReCS as part of the setup
for the spectroscopic observations. T-ReCS imaging was performed in the 11.7
$\mu$m narrow-band ($\Delta\lambda = 1.1\ \mu$m) filter, with an integration
time of 60 sec. Figure 1 displays the T-ReCS 11.7 \micron\ image.  At the
spatial resolution of T-ReCS, the binary system is clearly resolved with no
detection of extended emission connecting the system components; all of the
mid-infrared emission from SVS20-S and SVS20-N emanates from within $\sim$ 100
AU of the central protostars. Flux calibration was obtained from 11.7 $\mu$m
imaging of HD 187642. Standard NOAO IRAF packages were used to reduce the data
and extract the aperture photometry.

SVS20 was also imaged on 2002 July 21 (UT) with the Thermal Infrared Multimode
Instrument-2 (TIMMI2) on the ESO La Silla 3.6m telescope.  TIMMI2 has a $320
\times 240$ pixel Si:As blocked impurity band detector, with a pixel scale of
$0\farcs2$ pixel$^{-1}$. SVS20 was imaged in three narrowband filters
(N$\lambda$10.4, $\Delta\lambda=1\micron$; N$\lambda$11.9,
$\Delta\lambda=1.2\micron$; N$\lambda$12.9, $\Delta\lambda=1.2\micron$).  A
standard north-south chop-nod sequence with a 10\arcsec\ on-chip chop throw
was used. The integration time was 258 s. The data were reduced with
custom-written IDL routines for the TIMMI2 data format.  Standard aperture
photometry was performed using an IDL version of DAOPHOT.  Flux calibration
was obtained from observations of HD
187642\footnote{http://www.ls.eso.org/lasilla/sciops/3p6/timmi/html/stand.html}.

The T-ReCS spectra, along with the corresponding photometry from T-ReCS,
TIMMI2, and the literature, are summarized in Figure 2 and Table 1. In Figure
2,  the continuum levels for the mid-infrared emission underlying the observed
silicate features are estimated via a linear interpolation of the mean flux
density levels near the endpoints of the spectra: $\left [8.0 - 8.1\
\mu\rm{m}\right ]\ \&\ \left [12.8 - 12.9\ \mu\rm{m}\right ]$.  The T-ReCS
spectroscopic flux densities are in good agreement with the T-ReCS and TIMMI2
photometry and with the unresolved IRAS and ISO \citep[not shown,
see][]{alexander03} observations. The 10.78 $\mu$m flux densities for SVS20
reported by \citet{haisch02} are systematically lower by $\sim 25$\%. Given
the general agreement between the ISO, IRAS, T-ReCS, and TIMMI2 data, the 1998
flux densities reported by \citet{haisch02} may be discrepant or may represent
true intrinsic mid-infrared variability of SVS20 \citep[e.g.,][]{liu96}.

\section{Discussion}

The primary result of this paper is the visibly different mid-infrared spectra
exhibited by the two components of the binary system (Fig.~2).  Relative to
the continuum levels, SVS20-S displays a strong silicate emission spectrum,
while SVS20-N exhibits a shallow silicate absorption spectrum.
\citet{alexander03} modeled the {\em unresolved} ISOCAM CVF observations of
SVS20 as a combination of amorphous silicate absorption $(\tau \approx 0.6 -
0.8; A_V \sim 10 - 14$ mag) superposed upon amorphous silicate emission
$(\tau\approx -0.4)$. The T-ReCS spectra show that SVS20-S is responsible for
the majority of the observed silicate emission in SVS20.

\subsection{Component Luminosities}

Superficially, SVS20-S and SVS20-N have similar looking infrared spectral
energy distributions (SED) with $2.2 - 10$ \micron\ spectral indices $[ \alpha
= - d\ {\rm log}(\nu F_\nu)/d\ \rm{log}(\nu)]$ of $\alpha_{South} = 0.11 \pm
0.01$, $\alpha_{North} = 0.38 \pm 0.03$.  These spectral indices are redder
than those calculated by \citet{haisch02}, but still indicate that both
components are class I/flat spectrum protostars.

To explore the SEDs more carefully and parameterize the relative temperatures
and luminosities of the central protostars, we have fitted the near-infrared
($0.9 - 3.5$ \micron) photometry from the literature with a blackbody function
modified by a line-of-sight extinction curve: $S_\nu = \Omega B_\nu(T_d)\exp{(
{- {\rm A}_\nu}/1.086 ) }$, where $B_\nu(T)$ is the Planck function, ${\rm
A}_\nu$ is the frequency-dependent extinction \citep[R=3.1
assumed,][]{mathis90}, and $\Omega$ is the solid angle.

A temperature range of ${\rm T}=500-50000$ K in steps of 100 K and an
extinction range of ${\rm A_V}=0-50$ mag in steps of 0.05 mag were tested.
Uncertainties for the blackbody fitting were estimated via a Monte Carlo
simulation where the data points were randomly adjusted by their individual
uncertainties, and the data were re-fitted; the model parameter uncertainties
were estimated from the standard deviations of the best fits.

Of course, the SEDs arise from multiple emission sources of various
temperatures within the young stellar objects (e.g., central protostar,
accretion emission), but the near-infrared emission ($\lambda \lesssim 3\
\micron$) should be primarily from the stellar photospheres
\citep[e.g.,][]{gl97}, allowing us to estimate the relative temperatures and
luminosities of the central objects within each component. Results of the
fitting are shown in Figure 3. The SED for SVS20-N is reasonably well fit with
a ${\rm T}_{eff} = 3300\pm500$ K blackbody screened by an extinction of ${\rm
A_V} = 26.5\pm 1.5$ mag ($\chi^2_\nu \approx 2$) with a luminosity of
L$_\star\approx$ 0.9 L$_\odot$ -- similar to the values for other low-mass
young stellar objects \citep[e.g., IRAS 04016+2610: ${\rm T}_{eff} =
3300-4200,\ {\rm A_V} = 19-21$,][]{iti04}.

The SED modeling for SVS20-S is less constrained. The reduced chi-square for
the fits are never better than $\chi^2_\nu \approx 5$, regardless of the
combination of temperature and extinction.  There is a broad localized minimum
for the temperature range ${\rm T}_{eff} \approx 7000-10000$ K with
corresponding extinctions of ${\rm A_V} = 28-29$ mag. Models with lower
extinction have local minima at correspondingly lower temperatures, but the
fits are significantly worse ($\chi^2_\nu > 6$). Models with higher extinction
do not have localized minima; the reduced chi-squares asymptotically approach
$\chi^2_\nu \sim 5$. Figure 4 displays the dependence of the model chi-square
as a function of temperature for four different extinction levels. We
conservatively estimate that the central protostar for SVS20-S is a 10,000 K
blackbody, screened by $\sim 30$ mag of extinction -- suggesting a luminosity
of L$_\star\approx 20-80$ L$_\odot$, which is comparable to estimates made by
\citet{el87}.

The extinguished single-temperature blackbody fits to the near-infrared SEDs
represent estimates of the total visual extinction of the central protostar by
the surrounding circumstellar disk and the molecular envelope.  If a common
envelope contributes $\sim$14 magnitudes of extinction \citep{el87}, then the
circumstellar material immediately surrounding each protostar extinguishes the
central protostar by $10 - 15$ magnitudes. The modelling reduced chi-squares
of $\chi^2_\nu \sim 2-5$ are likely the result of the single-temperature
assumption, as evidenced by the extinguished blackbody models predicting too
little 10 \micron\ continuum emission (see Fig.~3).

\subsection{The Observed Mid-Infrared Spectra}

The observed mid-infrared spectrum of SVS20-S exhibits an emission feature
with the peak of the emission occurring near 11.3 $\mu$m.  In more evolved
sources such as HAeBe stars, the emission peak at 11.3 \micron\ is indicative
of crystalline silicate \citep[e.g.,][]{knacke93}, as amorphous olivine
particles ($0.1\ \mu$m in size) emit a feature peaking near 9.7 $\mu$m
\citep{bouwman01, ks05}. As amorphous silicate grains grow in size, the
emission feature broadens and becomes flat-topped \citep{bouwman01, prz03,
vanBoekel03}.

\citet{prz03} and \citet{vanBoekel03} correlate the strength of the silicate
emission feature with the ratio of the 11.3 $\mu$m to the 9.8 $\mu$m flux
densities.  A ratio of $F_{9.8\mu{\rm m}}/F_{11.3\mu{\rm m}} \lesssim 1$
suggests emission from crystalline silicates \citep{prz03}. \citet{ks05}
extend this technique by correlating the $F_{9.8\mu{\rm m}}/F_{11.3\mu{\rm
m}}$ and the $F_{9.8\mu{\rm m}}/F_{8.6\mu{\rm m}}$ ratios. They find that the
ratios are linearly correlated, and that sources with crystalline silicate
emission possess lower ratios than those sources with only amorphous silicate
emission.

For the observed SVS20-S spectrum, we have calculated the $F_{9.8\mu{\rm
m}}/F_{11.3\mu{\rm m}}$ and the $F_{9.8\mu{\rm m}}/F_{8.6\mu{\rm m}}$ ratios
from the continuum-normalized spectrum (Figure 5). We measured ratios of
$F_{9.8\mu{\rm m}}/F_{11.3\mu{\rm m}} = 0.8 \pm 0.1$ and $F_{9.8\mu{\rm
m}}/F_{8.6\mu{\rm m}} = 1.0 \pm 0.1$. These ratios place SVS20-S in the same
region of the $[8.6\micron, 9.8\micron, 11.3\micron]$ color-color diagram as
the more evolved T Tauri star Hen 3-600A and the HAeBe star HD179218, both of
which have crystalline silicate emission \citep[see Figure 11 in][]{ks05}.
However, these ratios for SVS20-S, while similar to those observed in more
evolved pre-main sequence stars with known crystalline grain emission, do not
take into account that SVS20-S, unlike the more evolved T Tauri and HAeBe
stars, has a significant amount ($\sim 14$ mag) of foreground absorption from
the surrounding envelope (see \S 3.3).

For SVS20-N, the observed mid-infrared spectrum, in contrast to SVS20-S, is
dominated by an absorption feature centered at 9.7 $\mu$m. The mid-infrared
spectrum of SVS20-N resembles other low-mass embedded class I young stellar
objects such as IRAS~04239+2436 \citep{ks05}, or IRAS~04108+2803B
\citep{watson04}, where the amorphous silicate absorption at 9.7 $\mu$m is
clearly present but shallow ($\sim 25\%$ below the continuum level). We have
calculated the optical depth of the absorption based upon the estimated
continuum level (Fig 2): $\ \tau_{\nu} = -\ln{(\frac{F_\nu}{F_{\nu o}})}$
where $F_\nu$ is the observed flux density and $F_{\nu o}$ is the flux density
of the continuum.  The optical depth for SVS20-N as a function of wavelength
is presented in Figure 5. Assuming ${\rm A_V}/\tau_{9.7\mu m} \approx 17$
\citep{rl85}, the peak optical depth of $\tau \sim 0.3$ corresponds to a
line-of-sight visual extinction of only ${\rm A_V} \sim 5$ mag.

The extinction derived from the silicate optical depth is nearly 10 magnitudes
lower than the envelope extinction implied by the 3.1 \micron\ ice feature
\citep{el87}. The lower observed optical depth implies that the mid-infrared
spectrum of SVS20-N contains silicate emission that is filling in partially
($\sim 75\%$) the line-of-sight absorption from the surrounding envelope. In
addition, there is a decrease in the optical depth near 11.3 \micron\
indicating that some of the emission may be from crystalline silicate
emission.   In the following section, the effect of the envelope absorption on
the innate mid-infrared emission from SVS20-S and SVS20-N is discussed.

\subsection{Correction for the Envelope Extinction}

To probe the effects of the line-of-sight absorption by the surrounding common
molecular envelope on the observed mid-infrared spectra, we developed a model
of the expected extinction by the dust grains contained within the common
envelope. Indices of refraction of amorphous olivine \citep{dorschner95} were
used with Mie theory to calculate the wavelength-dependent extinction
coefficients for particles 0.15 micron in diameter.  Based upon the 3.1
\micron\ ice feature \citep{el87}, we assumed that the peak line-of-sight
extinction produced by the envelope is A$_{\rm V}=14$ mag for both SVS20-S and
SVSV20-N. The envelope dust was assumed to be cool ($\sim 15$ K) enough to
produce no substantial mid-infrared emission of its own \citep{ts98}.  The
mid-infrared spectra corrected for the envelope extinction $[ F_{\nu o} =
F_{\nu}\exp\left ( \tau_{env} \right ) ]$ are presented in Figure 6. These
spectra represent the expected mid-infrared emission of the protostars, after
the removal of the surrounding common envelope extinction.

The protostellar mid-infrared emission is dominated by the emission of
amorphous silicates, which, in both protostars, peaks at $\lambda \approx 10$
\micron.  The silicate emission above the continuum from SVS20-S is four times
the strength of the silicate emission from SVS20-N.  This is likely a direct
result of the different temperatures and luminosities of the central objects
(\S 3.1), where SVS20-S is three times hotter and 20 times more luminous than
SVS20-N.  After the extinction correction, the $2.2 - 10$ \micron\ spectral
indices for SVS20-S and SVS20-N, respectively, are $\alpha \approx -0.3$ and
$\alpha \approx -0.1$, suggesting that while the binary is still deeply
embedded, the components of SVS20 may be evolving towards Class II young
stellar objects.

If we apply the flux density ratios developed by \citet{prz03},
\citet{vanBoekel03}, and \citet{ks05}, we find for SVS20-S, that
$F_{9.8\mu{\rm m}}/F_{11.3\mu{\rm m}} \approx 1.3\ {\rm and }\ F_{9.8\mu{\rm
m}}/F_{8.6\mu{\rm m}} \approx 2.0$, and for SVS20-N, that $F_{9.8\mu{\rm
m}}/F_{11.3\mu{\rm m}} \approx 1.3\ {\rm and }\ F_{9.8\mu{\rm
m}}/F_{8.6\mu{\rm m}} \approx 1.5$. These new ratios are significantly larger
than those derived prior to the correction of the line-of-sight extinction,
and taken by themselves place SVS20-S and SVS20-N among T Tauri and HAeBe
stars with no detectable crystalline emission. However, while the spectra for
both SVS20-S and SVS20-N are dominated by amorphous olivine emission, it is
evident from Fig. 6 that there is also additional emission near 11.3 \micron\
producing a shoulder on the broader olivine emission feature. To study this
feature in more detail, we have fitted a local continuum to the spectra
between $10.5 \leq \lambda \leq 12.1$ \micron. The continuum levels for the
mid-infrared emission underlying the observed features are estimated via a
quadratic interpolation of the flux density levels near the endpoints: $\left
[10.5 - 10.7\ \mu\rm{m}\right ]\ \&\ \left [11.8 - 12.1\ \mu\rm{m}\right ]$.
The continuum-subtracted spectra are displayed in the bottom panel of Figure
6.

The emission in this local region from SVS20-S remains four times the strength
of the emission from SVS20-N, but both sources show clear peaks near $\lambda
\approx 11.3$ \micron\ and $\lambda \approx 10.9$ \micron.  The peak centered
near 11.3 \micron\ tentatively is associated with forsterite Mg$_2$SiO$_4$
\citep{bouwman01}.  This forsterite peak is always accompanied by an
additional narrrow peak at 10 \micron\ which is evident in both sources (top,
Fig. 6).  The narrower peak at $\lambda \approx 10.9$ \micron\ tentatively is
associated with with enstatite MgSiO$_3$ \citep{bouwman01}. Enstatite is also
associated with broader peaks near $8.5 - 9.5$ \micron\ and 10.5 \micron.
Evidence for emission from these features in excess of the amorphous silicate
emission can also be seen in both sources (top, Fig. 6). Finally, as mentioned
in \S 2, there is no resolved extended mid-infrared emission indicating that
all of the silciate emission (amorphous and crystalline) is contained within
$\sim 100$ AU of the central protostars. Thus, both SVS20-S and SVS20-N appear
to have begun processing and annealing the dust grains within their local
environments.

\section{Summary}

We have presented spatially resolved mid-infrared ($8 - 13\ \mu$m) spectra of
the class I/flat-spectrum protostellar binary Serpens SVS20-(S,N). We
summarize our results as follows:

(1) Although SVS20-S and SVS20-N have similar broadband spectral energy
distributions, our simple models suggest that the temperatures and
luminosities of the two central objects are quite different. The central
protostellar source in SVS20-S has a effective temperature of T$_{eff} \approx
7000 - 10000$ K and a luminosity of L$_{\star} \approx 20 - 80$ L$_\odot$,
while the central protostar in SVS20-N has an effective temperature of
T$_{eff}\sim 3300$ K and a luminosity of only L$_{\star} \sim 0.9$ L$_\odot$.

(2) The observed mid-infrared spectra of the binary components also differ.
The spectrum of SVS20-S exhibits strong silicate emission, while that of
SVS20-N is dominated by shallow amorphous silicate absorption.  After
correction for the line-of-sight common envelope extinction, the mid-infrared
emission of the protostars for the two objects are dominated by amorphous
silicate emission, although the emission feature from SVS20-S is four times
the strength of the emission feature from SVS20-N. In addition to the
amorphous silicate emission, {\em both} SVS20-S and SVS20-N show evidence for
the presence of emission from crystalline silicate in the form of forsterite
and enstatite.

(3) The different luminosities of the two components may be responsible for
the different amounts of crystalline silicate emission observed in each
component. The lower luminosity of SVS20-N may have resulted in longer
processing times for the circumstellar material, although the apparent
presence of crystalline emission indicates that, even for lower mass young
stellar objects, crystallinization occurs relatively early in the star
formation process.

In order to understand more fully the binary system SVS20, modelling of the
complete spectral energy distribution coupled with accretion processes and
mineralogy will be the subject of a future paper. The differences in the
spectra clearly indicate a different level of dust processing between the two
protostars.  The presence of crystalline silicates in such a young binary
protostellar system indicates that the grain growth and processing found in
more evolved pre-main sequence stars, in debris disks around main sequence
stars, and in comets in our own solar system begins at a much younger
evolutionary stage while accretion is still ongoing.

\acknowledgments

The authors would like thank the engineering staff at the University of
Florida and the staff at Gemini Observatories for their help and outstanding
support in making T-ReCS a success.  The authors thank the referee whose
comments improved the paper greatly. The research was supported in part by NSF
grant AST-0098392 to C.M, and by NSF grant AST-0206617 to C.P. and J.T.R.
Based on observations obtained at the Gemini Observatory, which is operated by
the Association of Universities for Research in Astronomy, Inc., under a
cooperative agreement with the NSF on behalf of the Gemini partnership: the
National Science Foundation (United States), the Particle Physics and
Astronomy Research Council (United Kingdom), the National Research Council
(Canada), CONICYT (Chile), the Australian Research Council (Australia), CNPq
(Brazil) and CONICET (Argentina).

\newpage

\begin{deluxetable}{ccccc}
\tablecolumns{5}
\tablewidth{5in}
\tablecaption{Summary of Mid-Infrared Flux Densities}
\tablehead{
\colhead{$\lambda_c$} & \colhead{SVS20-S} &
\colhead{SVS20-N} & \colhead{Comments} & \colhead{Reference}\\
\colhead{($\mu$m)} & \colhead{$F_\nu$ (Jy)} & \colhead{$F_\nu$ (Jy)} & &
 }

\startdata

8.0 & $4.0\pm0.1$ & $2.6\pm0.1$ & T-ReCS Spectroscopy & 1\\
8.0 & \multicolumn{2}{c}{$6.16\pm0.14$} & ISOCAM CVF& 2\\
    &               &         & Binary Unresolved & \\
    &&&&\\
10.4 & $5.4\pm0.5$ & $2.5\pm0.2$ & TIMMI2 Photometry & 1\\
10.78 & $4.36\pm0.26$ & $1.53\pm0.09$ & MIRLIN Photometry & 3\\
    &&&&\\
11.7 & $5.32\pm0.27$ & $2.80\pm0.14$ & T-ReCS Photometry & 1\\
11.9 & $4.7\pm0.5$ & $3.0\pm0.3$ & TIMMI2 Photometry & 1\\
12.0 & \multicolumn{2}{c}{$8.5\pm0.6$} & IRAS/HIRES & 4\\
    &               &         & Binary Unresolved & \\
    &&&&\\
12.9 & $3.6\pm0.4$ & $3.2\pm0.3$ & TIMMI2 Photometry & 1\\

\enddata

\tablerefs{1. This work, 2. \citet{alexander03}, 3. \citet{haisch02}, 4.
\citet{hb96}}

\end{deluxetable}

\newpage

\begin{figure}
    \epsscale{0.75}

    \plotone{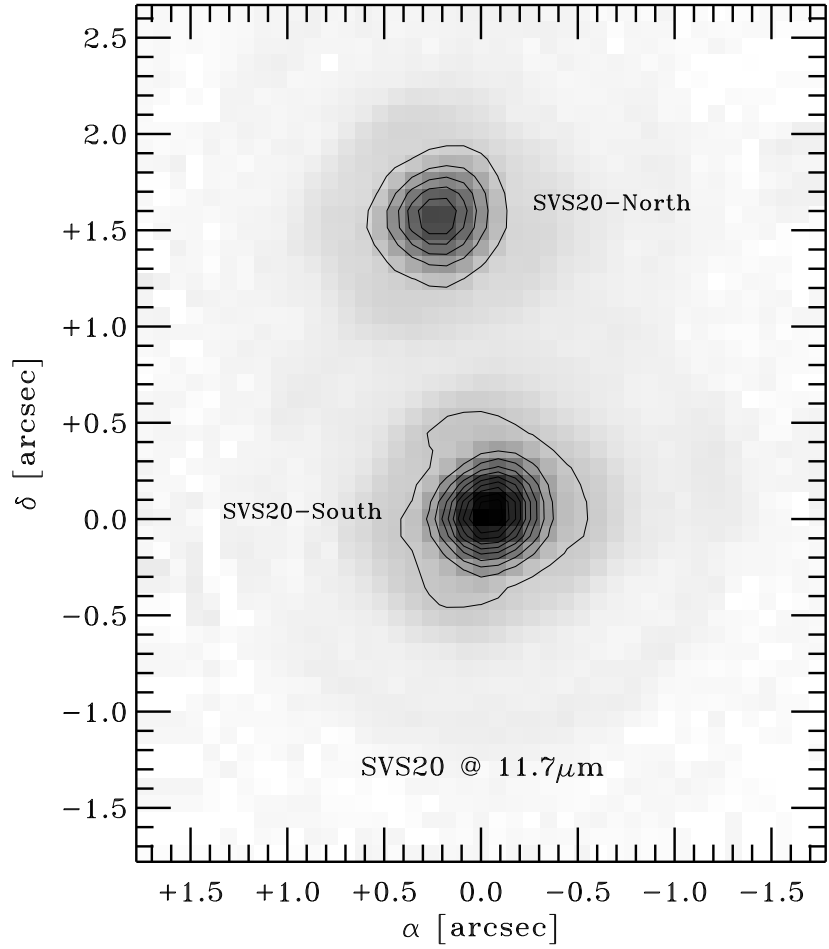}

    \figcaption{T-ReCS 11.7\micron\ image of SVS20.  The (0,0) point of the
    image is centered on SVS20-South at $\alpha=18h29m58.7s,\ \delta=
    01d14m03.2s\ (J2000)$.  The greyscale has been stretched by a square-root
    to enhance the contrast.  The contours are linear and start at 0.01 Jy
    pix$^{-1}$ and are stepped by 0.02 Jy pix$^{-1}$.}

\end{figure}

\begin{figure}
    \epsscale{0.75}

    \plotone{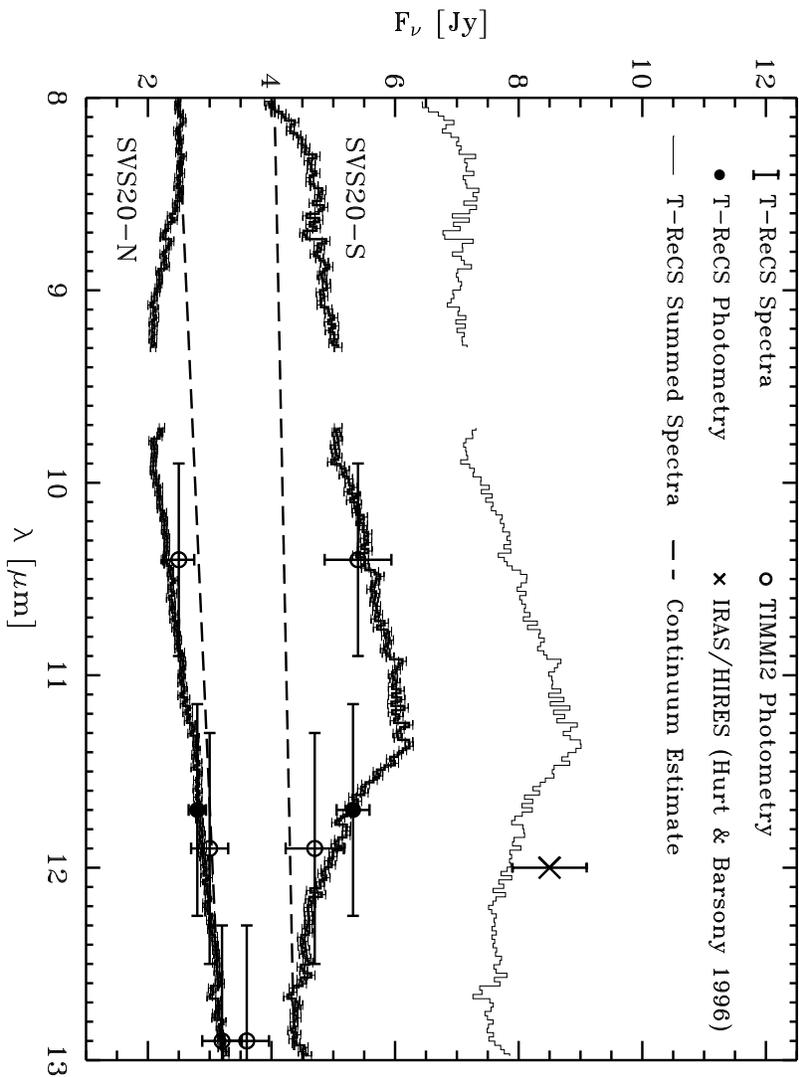}

    \figcaption{Mid-infrared spectra for SVS20-S and SVS20-N.  The individual
    data points represent photometry presented in this work (T-ReCS \& TIMMI2
    photometry) and from the literature (IRAS).  The summed spectrum is
    presented for ease of comparison to the unresolved IRAS photometry
    \citep{hb96} and the unresolved ISO spectra \citep[not shown;][]
    {alexander03}.  The data near telluric ozone ($9.3 < \lambda < 9.7\ \mu$m)
    have been removed because of uncertain ozone subtraction.  The abscissa
    error bars on the narrowband photometric points represent the widths
    of the narrowband filters.  The continua estimates to the spectra are shown
    as the dashed lines.}

\end{figure}

\begin{figure}
    \epsscale{0.75}

    \plotone{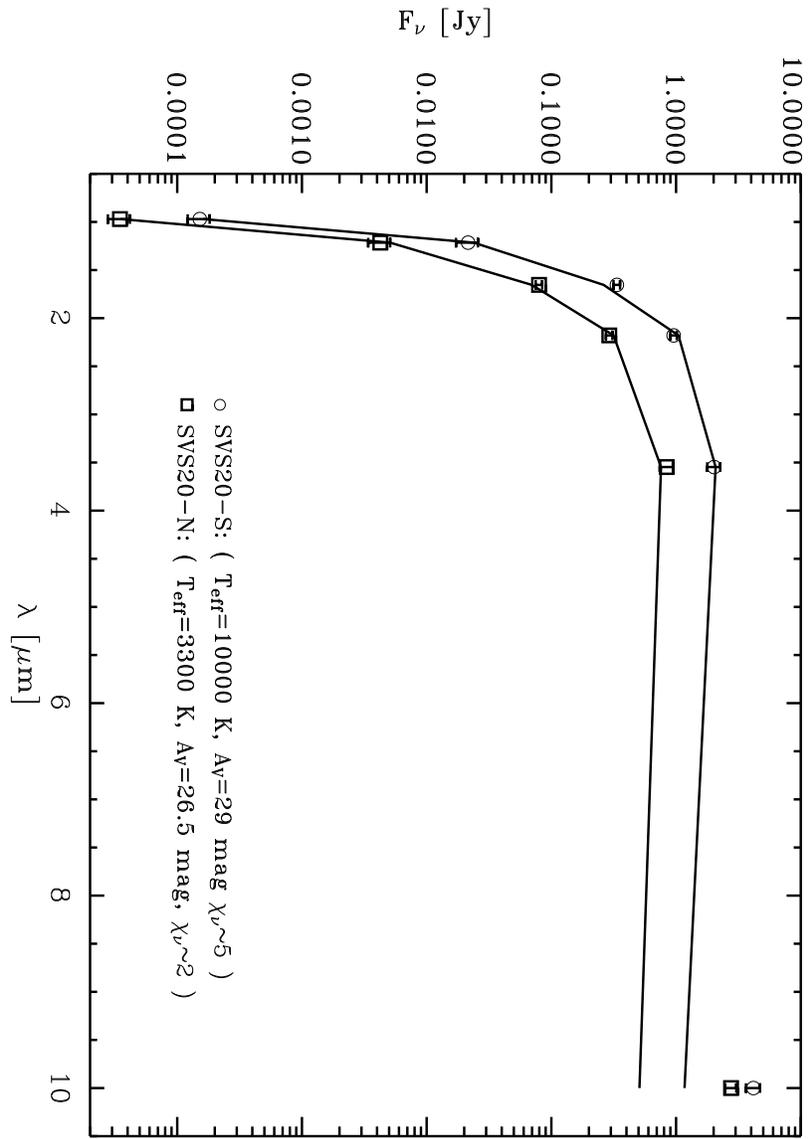}

    \figcaption{The spectral energy distributions for SVS20-S and SVS20-N. The
    models (solid lines) are extinguished single-temperature blackbody functions fit to the
    NIR data \citep{el87, ellh87, haisch02}.  The 10 \micron\ data, estimated
    from the {\em continuum} levels in Figure 2, are not included in the
    fitting.}

\end{figure}

\begin{figure}
    \epsscale{0.75}

    \plotone{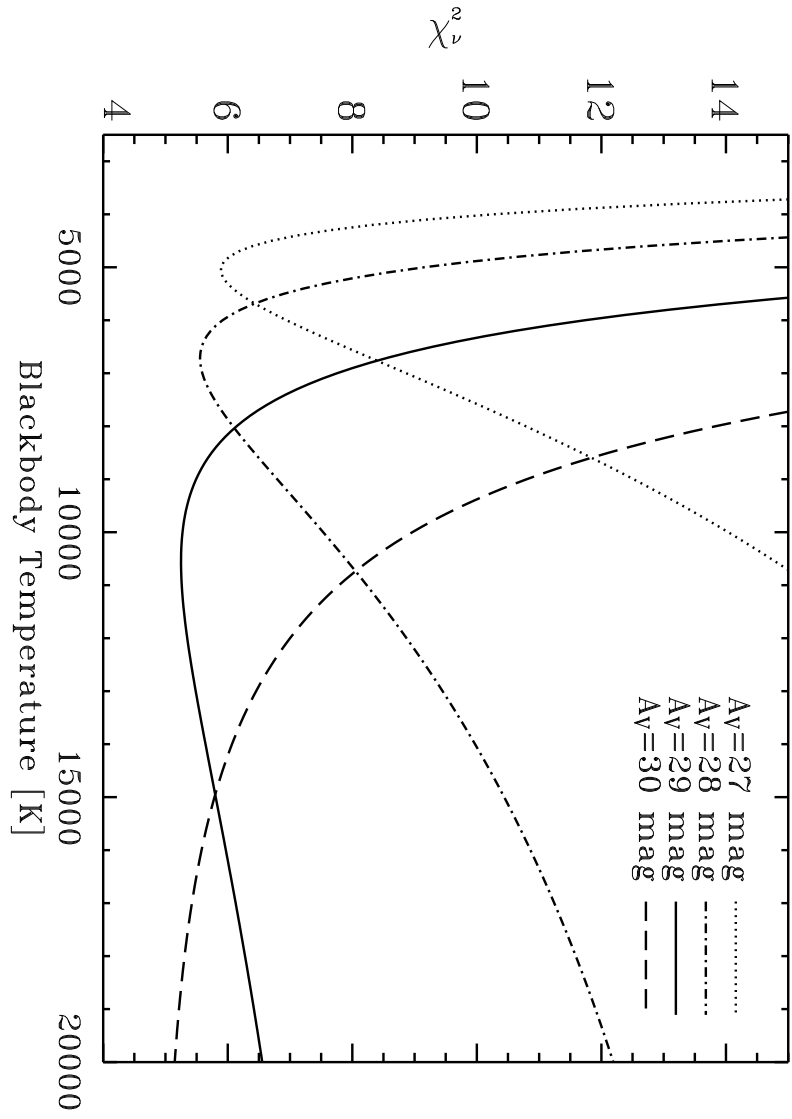}

    \figcaption{Plot of the reduced chi-square for the SED fits for SVS20-S as
    a function of temperature for four different visual extinction levels:
    A$_{\rm V} = 27$ mag, 28 mag, 29 mag, and 30 mag.}

\end{figure}

\begin{figure}
    \epsscale{0.75}

    \plotone{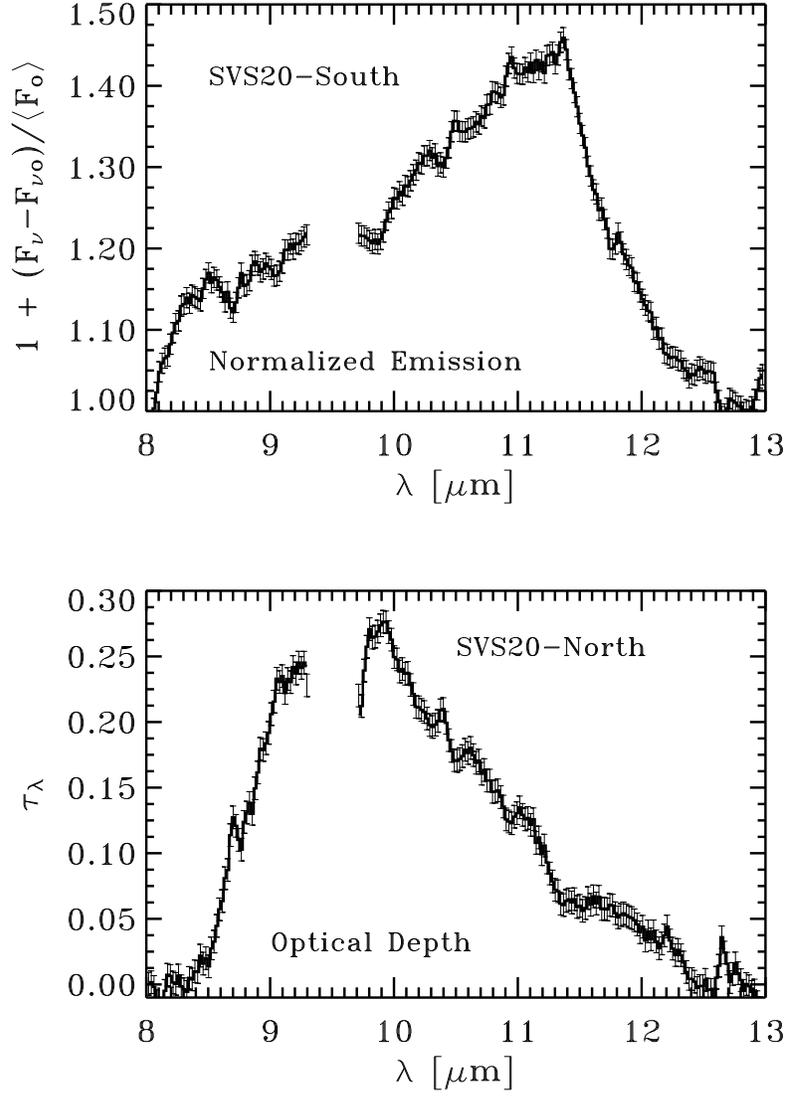}

    \figcaption{{\em Top:} Normalized emission of SVS20-South, median smoothed with
    a 3-pixel boxcar. {\em Bottom:}
    SVS20-North optical depth derived from the continuum-normalized spectrum, median
    smoothed with a 3-pixel boxcar.}

\end{figure}

\begin{figure}
    \epsscale{0.8}

    \plotone{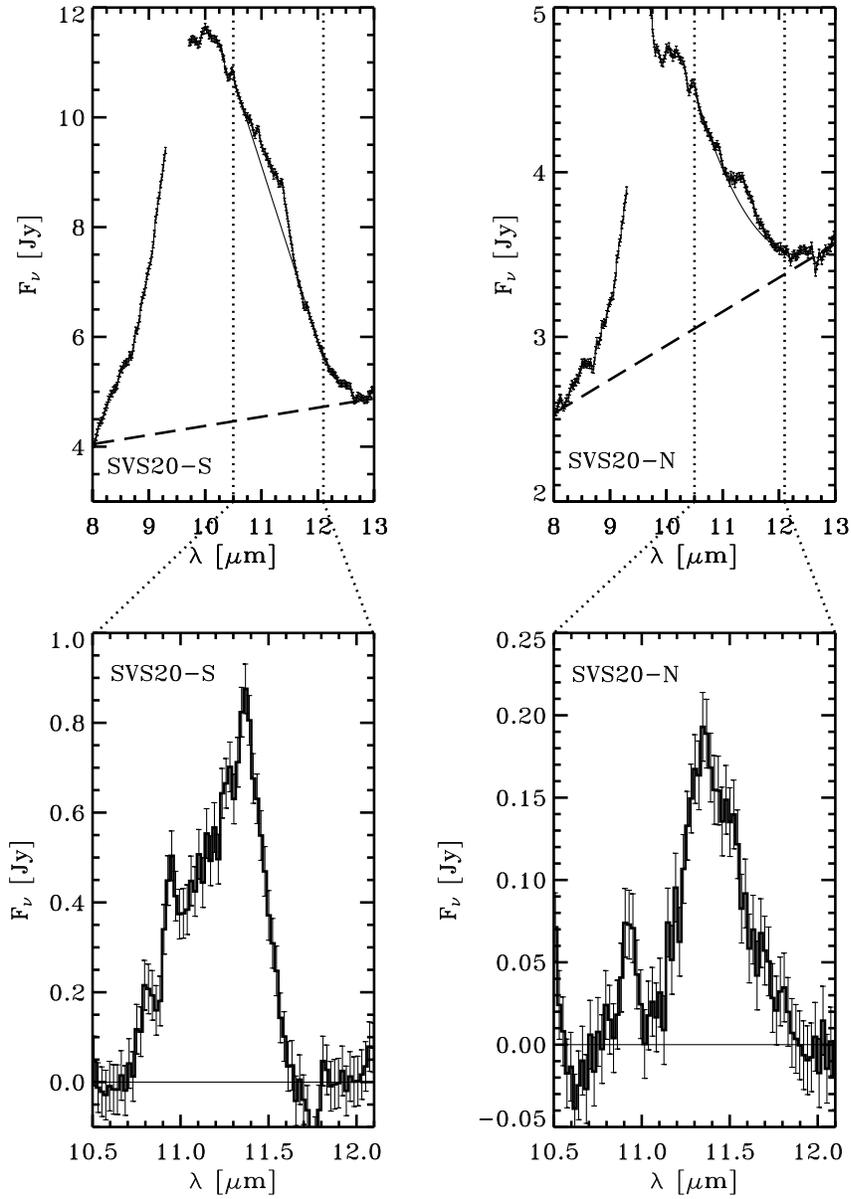}

    \figcaption{{\it Top:} Spectra for SVS20 after the removal of the envelope
    line-of-sight extinction model.  The dashed line represents the continuum
    level for the overall emission.  The dotted lines mark the wavelength range
    ($10.5 \leq \lambda \leq 12.1$ \micron) fitted with a local continuum
    represented by the solid lines.  Spectra have been median smoothed with a
    3-pixel boxcar. {\it Bottom:} The $10.5 \leq \lambda \leq 12.1$ \micron\
    spectra for SVS20-South and SVS20-North after the subtraction of the local
    continuum. Spectra have been median smoothed with a 3-pixel boxcar. }

\end{figure}

\end{document}